# Measurement: still a problem in standard quantum theory

R. E. Kastner
University of Maryland
August 21, 2013

ABSTRACT. It is argued that recent claims by A. Hobson that standard quantum theory has no measurement problem cannot be sustained. Moreover, it is pointed out that taking the reduced density operator of a component system as an epistemic (ignorance-based) mixture has long been known to be untenable based on direct proof.

A. Hobson (2013a,b) has recently argued that there is no measurement problem in quantum mechanics. He sums up his objections to those who assert that there continues to be a measurement problem by saying 'The primary fallacy …is that they fail to take entanglement and nonlocality into account.' But in fact, entanglement and nonlocality do no work whatsoever in bringing about the local collapse asserted by the author.

The author invokes decoherence as the mechanism providing determinacy of the basis of the local measurement. While even this claim can be questioned (see , e.g., Fields 2011), I leave it aside for the moment to focus on two arguments by the author in support of his claim that there is no measurement problem (i.e., no 'problem of outcomes').  The first of these I will call the 'no larger system superposition' argument  (1), or 'NLSS', and the second the ' improper mixtures are epistemic' argument (2), or 'IME'.

Hobson considers a composite system of S (microscopic system under study)  and A (measuring apparatus) in the pure entangled state

$$|\psi\rangle_{SA} = c_1|s_1\rangle|a_1\rangle + c_2|s_2\rangle|a_2\rangle,$$

which he labels equation (2).

Let's first consider 'NLSS.'  The author says that "the measurement problem arises from an oversimplification, namely the notion that $|\psi\rangle_{SA}$ can be thought of as a simple superposition $c_1|b_1\rangle + c_2|b_2\rangle$ of a larger system $B = SA$, where $|b_i\rangle = |s_i\rangle|a_i\rangle$ $(i=1,2)$. If S is a nucleus and A is Schrodinger's cat, this appears to be a superposition of an alive and dead cat--an absurd conclusion. Furthermore, such a superposition contradicts the collapse postulate of standard quantum physics--the assumption that, following the measurement, A will be in one of the states $|a_i\rangle$ and S will be in the corresponding state $|s_i\rangle$, with probabilities $p_i = |c_i|^2$. "

But in fact it is perfectly legitimate to take $|\psi\rangle_{SA}$ as a superposition of a system with two degrees of freedom  $|s_i\rangle|a_i\rangle$. For example, an atom of orthohelium has two electrons in a superposition of states of parallel spin.  Indeed, an appropriate measurement could find that the two electrons were both 'up' with a probability $|c_{up}|^2$. So there is no oversimplification involved here. The author then notes that an 'alive and dead' cat is an absurdity, which is simply to restate the measurement problem; i.e., stating that the measurement problem confronts us with an absurdity is not to show that it is solved. He then asserts that taking the state (2) as a superposed state of two systems (i.e. two degrees of freedom) contradicts the *assumption* (i.e., the collapse postulate) that measurement of one of them causes it (somehow) to collapse -- which is again simply to repeat the measurement problem, not to show that it is solved.

The collapse postulate is *an ad hoc* postulate, based on the *observational* fact that we don't observe superpositions, and it is the fact requiring explanation. The author is therefore apparently arguing that there is no measurement problem because *we see collapsed states* and because of a collapse postulate, put in by hand to accommodate this empirical fact. This is circular reasoning.

The author's second argument, 'IME,' concerns the reduced state of a component system, which is an improper mixture. The author wishes to apply an epistemic interpretation to this reduced state – i.e., he wants to argue that S is locally collapsed, which means that it really must be in one local state or another. However, R. I. G. Hughes (1992) provides a well-known proof that given a composite system in a pure entangled state, applying an epistemic interpretation to the improper mixed state of a component system fails. Hughes shows that taking the improper mixture as representing ignorance of which state the subsystem is actually in results in a contradiction: i.e., it implies that the composite system is in a mixed state, contrary to the assumed composite pure state.

Now, the author apparently wants to use the fact that the subsystem S is not in a *single-space* superposition to argue that it is locally collapsed. But as Hughes shows, if S approaches its local measuring device in a collapsed state, describable by an epistemic (proper) mixed state, that contradicts the prepared composite pure state. Yet the author appears to be asserting that the entangled composite state (2) really does represent locally collapsed component states of this type, since he says:

"The second theoretical objection is that the reduced density operators are "improper," because they arise not from ignorance about the quantum states of *S* and *A* but rather by reduction from the composite pure state $|\psi\rangle_{SA}$--the state that *S* and *A* are said to be "really" in. But this ignores the global versus local distinction that is central to resolving the measurement problem. Unlike a global observer, local observers at *S* and *A must be* uncertain about the (local) quantum states of these systems in order to preserve Einstein causality, and thus Eqs. (4) [the improper mixed states of the component systems] are the physically appropriate density operators. *S and A are really in their local states*, not the global state, because these are the states we always directly observe; the global state merely conveys the correlations between local states, correlations that cannot be observed locally without violating Einstein causality. … We've seen that, despite superficial appearances, Eq. (2) is the collapsed state, exhibiting definite outcomes for both sub-systems." (my italics)

There are a number of serious problems with these comments. First, the author seems to be questioning whether the composite system 'really' is in the state (2). But clearly, it is. Then he notes that a local observer must be uncertain about the local state of his system, based on a requirement to preserve Einstein causality. Of course! -- but an observer can readily be uncertain about the local state of his system *if that local state really is* indeterminate -- so the Einstein causality requirement is already preserved through the improper mixture itself. Put differently, the author seems to be invoking an epistemic mixture as a *necessary* condition for preserving Einstein causality when it is merely *sufficient*; but the improper mixture is certainly sufficient, so an epistemic mixture is not required. So once again, the improper, non-epistemic mixed states are indeed the physically appropriate density operators for the component systems. Then the author simply asserts, by fiat, that the local systems should be described by *epistemic* (collapsed) mixed states because 'these are the states we always directly observe' – which is again to *invoke* observation to *explain* an observation. That is, the whole point of the measurement problem is that the *theory* cannot explain why we always find collapsed states. Morover, in view of the proof by Hughes, one cannot take the improper mixed states (4) as epistemic, since this contradicts the state (2).[1]

The core of the measurement problem is the inability of the standard theory to explain the transformation from the linear unitary evolution (von Neumann's 'process 2') to the non-unitary evolution (von Neumann's 'process 1') of the collapse; i.e., the transformation from a pure state to a *proper* mixed state. Hobson invokes the collapse postulate and observation, but provides no physical explanation for this transformation, and therefore fails to solve the measurement problem. He seems to be confusing his *demand* that the theory yield collapse with a demonstration of collapse on the theoretical level that never is actually achieved – instead, it is noted that a cat in a superposition is an absurdity, and that we always observe collapsed phenomena , which is simply to repeat the measurement problem rather than to solve it. Thus, Weinberg (2012) is completely correct that in the usual theory, "during measurement the state vector of the microscopic system collapses in a probabilistic way to one of a number of classical states, in a way that is unexplained and cannot be described by the time-dependent Schrodinger equation."

However, a clear physical account of the transformation from the unitary evolution to the non-unitary 'collapse' process can be given in a direct-action picture, when the response of absorbers is taken into account (Kastner 2012, Chapter 3). Indeed, Von Neumann's 'process 1,' including the irreversibility of measurement, falls directly out of this analysis. I suggest that this is the methodologically correct way to solve the measurement problem -- from a specific physical process not accounted for by the usual theory, rather than by *invoking* observation to try to *explain* the observation that we see collapsed states when we perform measurements.

Acknowledgment. I am grateful to A. Hobson for correspondence.

---

[1] I should note that Kirkpatrick (2001) has criticized the Hughes' proof and a similar argument by d'Espagnat (1999); d'Espagnat (2001) has refuted these criticisms by demonstrating that considering each component system as 'really in' a given collapsed state must yield probabilities inconsistent with the pure state in which the composite system was prepared.